\documentclass[aos,preprint]{imsart}
\usepackage[]{graphicx}
\usepackage[]{color}
\usepackage{xcolor}

\makeatletter
\def\maxwidth{ %
  \ifdim\Gin@nat@width>\linewidth
    \linewidth
  \else
    \Gin@nat@width
  \fi
}
\makeatother

\usepackage{framed}
\makeatletter
 {\par\unskip\endMakeFramed%
 \at@end@of@kframe}
\makeatother

\definecolor{shadecolor}{rgb}{.97, .97, .97}
\definecolor{messagecolor}{rgb}{0, 0, 0}
\definecolor{warningcolor}{rgb}{1, 0, 1}
\definecolor{errorcolor}{rgb}{1, 0, 0}

\usepackage{alltt} 
\usepackage[latin1]{inputenc}
\usepackage{enumitem}
\usepackage{todonotes}
\usepackage{latexsym}
\usepackage{amssymb}
\usepackage{epsfig}
\usepackage{amsmath}
\usepackage{xcolor}
\usepackage{float}
\usepackage{hyperref}
\usepackage{booktabs}
\usepackage{longtable}

\allowdisplaybreaks

\IfFileExists{upquote.sty}{\usepackage{upquote}}{}

\usepackage{algorithm} 
\usepackage{algpseudocode} 
\makeatletter
\newcommand{\algmargin}{\the\ALG@thistlm}
\makeatother
\newlength{\whilewidth}
\algnewcommand{\parState}[1]{\State%
	\parbox[t]{\dimexpr\linewidth-\algmargin}{\strut #1\strut}}
\usepackage[latin1]{inputenc}
\usepackage{enumitem}
\usepackage{todonotes}
\RequirePackage[OT1]{fontenc}
\RequirePackage{amsmath,amssymb,amsthm}
\RequirePackage[colorlinks,citecolor=blue,urlcolor=blue]{hyperref}
\usepackage{graphicx}
\usepackage[11pt]{moresize}

\startlocaldefs
\numberwithin{equation}{section}
\theoremstyle{plain}

\endlocaldefs

\theoremstyle{plain}
\long\def\comment#1{}

\theoremstyle{definition}

\numberwithin{definition}{section}

\numberwithin{remark}{section}

\renewcommand{\qed}{\hfill{\tiny \ensuremath{\blacksquare} }}%

\begin{document}

\begin{frontmatter}
\title{Machine Learning for Financial Forecasting, Planning and Analysis: Recent Developments and Pitfalls}
\runtitle{Machine Learning for FP\&A}

\begin{aug}
\author{\fnms{Helmut} \snm{Wasserbacher*}\ead[label=e1]{}}
\and
\author{\fnms{Martin} \snm{Spindler}\ead[label=e2]{}}
\thankstext{T1}{* The views and opinions expressed in this document are those of the first author and do not necessarily reflect the official policy or position of Novartis or any of its officers.}


\address{Helmut Wasserbacher\\
Novartis International AG\\
Novartis Campus\\
4002 Basel\\
Switzerland\\
E-mail: helmut.wasserbacher@novartis.com}

\address{Martin Spindler\\
University of Hamburg\\
Hamburg Business School\\
Moorweidenstr. 18\\
20148 Hamburg\\
Germany\\
E-mail: martin.spindler@uni-hamburg.de}
\end{aug}

\begin{abstract}
This article is an introduction to machine learning for financial forecasting, planning and analysis (FP\&A). Machine learning appears well suited to support FP\&A with the highly automated extraction of information from large amounts of data. However, because most traditional machine learning techniques focus on forecasting (prediction), we discuss the particular care that must be taken to avoid the pitfalls of using them for planning and resource allocation (causal inference). While the naive application of machine learning usually fails in this context, the recently developed double machine learning framework can address causal questions of interest. We review the current literature on machine learning in FP\&A and illustrate in a simulation study how machine learning can be used for both forecasting and planning. We also investigate how forecasting and planning improve as the number of data points increases.

\end{abstract}



\end{frontmatter}


\section{Introduction}

Accurate financial forecasts and plans for effective and efficient resource allocation are core deliverables of the finance function in modern companies. Particularly in volatile or fast evolving market environments, fast and reliable forecasting and planning are crucial \cite{Becker2016}. High-quality forecasting is among the defining characteristics of strong finance functions \cite{Roos2020}. It is therefore hardly surprising that most larger companies have dedicated teams for financial planning and analysis (FP\&A) within their finance function.\\

The increasing availability of big data, coupled with new analysis techniques, provides an opportunity for FP\&A to generate more and better insights at a faster pace, generating more value for the company. Machine learning is a set of techniques developed in computer science and statistics that appear particularly well suited to this context. The aim of our paper is to show how machine learning can be used for FP\&A and which pitfalls can arise in the process. Machine learning has been applied successfully to a variety of predictive tasks, including fraud detection and financial forecasting. Planning and resource allocation, however, represent tasks of a different nature because they require understanding the effect of an active intervention in a system, such as the market for a product. For this reason, they are causal problems, which are harder to model with machine learning. A large field within machine learning revolves around pattern recognition. Patterns in data, based on correlations, are learned and then used for predictions. In causal tasks, an understanding of the underlying (causal) mechanisms is important when evaluating the effects of interventions (e.g., the implementation of a new business strategy). The emerging field of causal machine learning uses machine learning algorithms for such questions. For instance, the recently developed double machine learning framework reduces the impact of imperfect model specifications, which are hard to avoid in practice in the context of causal analysis.\\

We structure this paper as follows. In Section~\ref{Role}, we briefly review the role of FP\&A. Section~\ref{ML} provides a short, focused overview of machine learning. In particular, we highlight the pitfall of not distinguishing between forecasting and planning. In Section~\ref{Literature} we present the results of our literature review, which finds surprisingly few publications of machine learning applications in FP\&A. In Section~\ref{Simulation}, we describe and provide the results from a simulation study. We compare a classical machine learning technique, the lasso, to a linear regression based on the ordinary least squares (OLS) method. In our analysis, we refer back to the distinction between forecasting and planning from Section~\ref{ML} and show how the results differ between the lasso and OLS for both tasks. Finally, we also quantify the benefit of additional data in this simulation.\\

\section{The role of FP\&A}
\label{Role}

Given the importance of financial forecasting, planning and analysis (FP\&A) in modern corporations, most larger companies have dedicated teams for these tasks within their finance function, even though the exact organizational design and naming of the department may vary.\footnote{In some companies, the (short-term) plans formally expressed in budgets are prepared by controllers (management accountants) within the accounting department \cite{Garrison2006}, while the strategy department formulates the directional (long-term) plans.}
The overarching goal of FP\&A is to inform and support decisions of management and the board of directors \cite{Oesterreich2019}. FP\&A pursues this goal via different routes, helping determine which projects in the company portfolio create value (and are consequently worth funding), and preparing company-wide forecasts and financial plans to ensure that the company can reach its financial goals in the short and long term \cite{Ross2019}. Investments in research and development (R\&D) or the expansion of production capacity are balanced with financial obligations to debt holders or equity investors and tax authorities \cite{Brealey2020}. Financial plans are also an important step in the translation of a company's strategic priorities into concrete operational actions. These actions contribute to focusing the organization and the deployed resources behind common goals.\\

Analyzing the business environment and business dynamics is an integral part of the work performed by FP\&A. The insights generated through such analysis can inform the development of forecasts and plans and help in the assessment of how likely these plans are to succeed. During the the execution phase of plans, these insights allow FP\&A to understand why actual results may deviate from the plan and to recommend corrective actions. This need for business acumen is likely to continue, even when advanced forecasting methods like those described in this article are used \cite{Moeller2020}.\\

The time horizons considered for financial forecasts and plans usually range from one month to several years \cite{Ross2019, Fischer2009}. The choice of time horizon depends on company-specific circumstances and objectives. For instance, stock-market listed companies typically put additional weight on quarterly figures. In practice, most companies create forecasts and plans for the next fiscal year (sometimes called a budget), which additionally can serve as a management control mechanism \cite{Strauss2013}. Rolling forecasts are another form of plan. These are characterized by regular updates, which are typically performed on a monthly or quarterly basis \cite{Hansen2011}.\\

FP\&A relies in large part on quantitative analysis to generate forecasts and plans. Accounting systems are a major internal data source for FP\&A \cite{Garrison2006, Gray2015}, covering items related to sales (turnover), expenses and balance sheet positions, which are especially important for cash flow analysis. Other important internal sources of data include those related to human resources (employee numbers, wage costs), supply chain and production (manufacturing costs at various levels of granularity) and R\&D (product development costs, success rates, timelines).\\ 

External data sources include market- or product-specific information, such as the size and development of the market and market shares. The exact nature and granularity of these data depends largely on the product or question under analysis, as well as the investment required to access the relevant data \cite{Gray2015}. For instance, it is not uncommon in the consumer goods industry to have access to transaction-level data \cite{Taddy2019}, covering one's own and competitor products. However, information at this level of specificity is typically used by the marketing and sales department for product-specific tactics. In contrast, FP\&A often uses macroeconomic indicators, including GDP, inflation and currency rates.\\

The development and spread of comprehensive, company-wide IT systems in recent decades has increased the amount and variety of data readily available to FP\&A. Increasing digitalization will further accentuate this development, with big data as the crystallizing term. The ``Three V's'', a common framework to define big data \cite{Laney2001}, allow us to look at the different dimensions that drive this development. First, the amount of information generated, captured and thus accessible for FP\&A activities is growing (volume). Second, the speed of information creation and its accessibility is accelerating (velocity); as a consequence, the speed at which information must be analyzed and acted upon increases, too. This calls for automated, real-time analytics and evidence-based planning \cite{Gandomi2015}. Third, more and more types of information are being gathered or generated and can be analyzed (variety); for instance, stock market analysts apply sentiment analysis to extract information relevant to stock prices from text documents.\\

In addition, other dimensions of big data have been proposed \cite{Gandomi2015}. In the context of FP\&A, the additional dimensions of veracity and complexity appear especially relevant. Thanks to the more widespread use of digital tools, the need for data transparency and scrutiny within many companies is increasing as well. In turn, the need to ensure data quality and reliability is growing (veracity). Moreover, (big) data are generated through multiple sources, both from inside and outside the company. This requires data cleaning, data matching and, ideally centralized storage, which facilitates accessibility (complexity).\\

As mentioned above, a key output of FP\&A is financial forecasts and plans. For data that are more numerous, available more quickly, and are more diverse and of better quality than in the past, FP\&A needs to choose adequate tools, such as those provided by machine learning.\\

\section{Introduction to machine learning}
\label{ML}

While there is no uniform definition of machine learning, it can be described as a collection of methods that automatically build  predictions from complex data \cite{Taddy2019}. In essence, machine learning deploys a function-fitting strategy aiming to find a useful approximation of the function that underlies the predictive relationship between input and output data \cite{Hastie2009}. In this search for patterns in data \cite{Bishop2006}, which, to a large extent, is executed autonomously, machine learning draws on statistical tools and algorithmic approaches from computer science. In particular, machine learning aims to cope with the situation of high-dimensional data. High dimensionality occurs when the number of input variables (independent variables, features) used to predict the output (dependent) variable is large compared to the number of observations available. Classical statistical techniques do not work in this setting \cite{Taddy2019}.\\

The three broad categories of machine learning are supervised learning, unsupervised learning and reinforcement learning. \textit{Supervised learning} is concerned with predicting the value of an output variable based on the values of a set of input variables. For this, supervised learning relies on a set of input and output variables that are jointly observed for each data point \cite{Hastie2009}. A practical example is to predict the sales of a product using input variables such as time of the year, price level, advertising expenditures and availability of competitor products. In contrast, \textit{unsupervised learning} consists only of a set of input observations for which the joint distribution is known. However, there is no observed output (response). The goal is to directly infer the properties of these observations \cite{Hastie2009}. Classifying customers into (previously unknown) customer archetypes based on their observed characteristics such as buying behavior, age, gender and socio-economic status is an example of unsupervised learning. In \textit{reinforcement learning}, the algorithm performs a trial-and-error search to maximize a numeric reward signal, in direct interaction with its environment \cite{Sutton2018}. By interacting with its environment, the algorithm creates its own data from which it can learn. Games such as checkers, chess and go are classical examples in which reinforcement learning is applied. Sometimes cited as a fourth category, \textit{semi-supervised learning} falls between supervised and unsupervised learning, combining a small amount of fully labeled data as in supervised learning and a large amount of unlabeled data as in unsupervised learning. The objective is to improve supervised learning in situations in which labeled data are scarce \cite{Zhu2009}. For the purposes of FP\&A objectives, which mostly revolve around producing forecasts from a set of inputs and assumptions, the predominant choice is typically supervised methods.\\

Machine learning methods appear especially suitable for the core FP\&A task of forecasting because of their focus on predictive performance. These methods manage to identify generalizable patterns that work well on new data, i.e., data outside of the training sample \cite{Mullainathan2017}. Through their ability to identify complex structures that have not been specified in advance, they lend themselves to support a high degree of automation in the generation of forecasts. This flexibility has the additional advantage that many off-the-shelf algorithms perform surprisingly well on a variety of tasks. In addition, a large selection of machine learning algorithms are available and are technically easy to use \cite{Mullainathan2017}, making them attractive for practitioners.\\

Besides forecasting, the second core task of FP\&A is to provide recommendations for the design of financial plans and for potential corrective actions when deviations from plans occur. In statistical terminology, this requires causal inference techniques, which are fundamentally different from forecasting. Consider the trivial example of hotel occupancy rates and room prices \cite{Athey2018}. High room prices coincide with high occupancy rates. Thus, price variations are strongly predictive of hotel occupancy. If the goal is to make a forecast, we do not need to be concerned with understanding why occupancy was high. However, if we want to recommend an action to increase the occupancy rate (an intervention) or imagine in retrospect what the occupancy rate would have been if the room rates had been different (``counterfactual'' \cite{Pearl2018} or ``potential outcome'' \cite{Rubin2005}), FP\&A requires a causal understanding of the business dynamics. To conclude with this example, a plan consisting of a room price increase will not lead to higher occupancy. Most likely, prices have increased in the past in reaction to high demand, which was stimulated by other factors (e.g., the holiday season). While this trivial example seems obvious, it illustrates a major pitfall: many companies struggle in practice to identify truly causal measures for the effectiveness of their promotional activities. \cite{Blake2015} discuss this phenomenon in the context of large-scale field experiments conducted at the e-commerce platform eBay.\\

For interventional and counterfactual analysis, data-driven approaches need to produce reliable estimates for the parameters that govern the relationship between input and output variables. Machine learning algorithms are typically not built for this purpose. Historically, the machine learning community has pursued the goal of maximizing predictive performance as opposed to understanding model parameters \cite{Taddy2019}. However, using a tool built for forecasting and assuming that it also possesses the properties required for causal inference in economic applications can be misleading \cite{Mullainathan2017}. Maximizing the predictive power of a model to use it for interventional analysis represents a major trap. Indeed, it may even be necessary to sacrifice predictive accuracy to arrive at a correct understanding of the relationships that are relevant for making decisions about interventions \cite{Athey2018}. The current lack of understanding of cause-effect connections is even cited as a fundamental obstacle for machine learning by some authors \cite{Pearl2019}. Nevertheless, many inference procedures include prediction tasks as an important step \cite{Mullainathan2017}. Machine learning is especially suited for this step in high-dimensional settings \cite{Belloni2014}. The double machine learning framework \cite{Chernozhukov2017}, which we will apply in Section~\ref{Simulation}, allows us to take advantage of the predictive performance of machine learning algorithms when seeking solutions for for causal problems.\\

\section{Literature review}
\label{Literature}


We conducted a search of the literature across Google, Google Scholar and finance journals on the use of machine learning in FP\&A. The use of quantitative methods in the broad field of finance has been studied intensively for close to 40 years \cite{Ozbayoglu2020}, in part because of the general availability of data in this field, the existence of many areas of implementation and the substantial economic impact of financial decisions. Our search yielded surprisingly few recent publications on the use of machine learning explicitly in FP\&A and related fields. The key thrust of machine learning in finance is directed towards various applications ultimately linked to forecasting and trading financial instruments such as stocks, bonds, currencies and derivatives. Credit scoring and fraud detection are other major areas. Examples of recent surveys include \cite{Ozbayoglu2020} and \cite{Henrique2019}.\\

We see two possible reasons for the apparent scarcity of publications on machine learning in FP\&A. First, time-series forecasting has been thoroughly covered and researched for many years \cite{DeGooijer2006}. A large variety of tools for this purpose have been developed, both from an academic and theoretical perspective, as well as from the perspective of practitioners, including easy to use off-the-shelf software \cite{Kuesters2006}. From a practical FP\&A perspective, these tools, together with the domain knowledge of the experts working in the FP\&A function, allow practitioners to arrive at results that - by and large - serve sufficiently well to meet the objective of developing financial plans. Especially practitioners may therefore perceive machine learning as a ``so-so'' technology \cite{Acemoglu2018}, which is not (yet) quite worth their (full) attention. Thus, the intrinsic urge to look for new tools, including machine learning, in FP\&A is still less pronounced than it is, for instance, in stock market forecasting, where even a relatively small improvement in forecasting accuracy can yield significant economic payoff. We believe that this will change with the further deployment of digitalization and the consequent increase in data availability as described above. Besides improving the precision of financial forecasts, automated forecasts driven by machine learning can also lead to a substantial reduction in costs and to increased flexibility given that the traditional process is quite labor- and time-intensive.\\

Second, we hypothesize the following reason for the limited number of publications on machine learning in FP\&A. The initial development of artificial intelligence and machine learning methods was driven mostly by academia. Because these methods are highly relevant for industrial applications, companies (in particular in the tech field) have shown strong interest in applying and developing them further. Indeed, some of the large tech companies host their own dedicated research teams. However, the limited availability of skilled professionals represents a hurdle to fast diffusion in all corporate functions of a company. Therefore, the application of machine learning for FP\&A is still rare in the finance function, even though a host of machine learning publications by the industry have already appeared in other functional areas.\footnote{For instance, \url{www.bosch.com/de/forschung/know-how/publikationen} (accessed Feb 23, 2021) contains a collection by Robert Bosch GmbH.} Management consultancies have also discovered the benefit of machine learning for finance and FP\&A. However, their publications remain general and directional in nature (see for instance \cite{Balakrishnan2020}, \cite{Roos2020}, \cite{Tucker2017}, \cite{Chandra2018}).\\

One company that has made public its use of machine learning in FP\&A in scientific papers is Microsoft Corporation. In the past several years, Microsoft appears to have followed an innovative approach with machine learning in FP\&A as witnessed by three publications from its employees. One paper \cite{Gajewar2016} compares the performance of random forests to that of traditional time-series methods such as autoregressive integrated moving average (ARIMA), error trend and seasonality (ETS, a variant of exponential smoothing) and seasonal-trend decomposition using loess (STL, another variant of smoothing) for forecasting quarterly revenues by major geographic region and at the global level up to one year into the future. Based on their exploratory analysis, the random forest model with a restricted number of features outperformed the traditional time series methods and the forecasts generated by the domain experts in the Microsoft FP\&A department.\\

A second paper \cite{Barker2018} describes a machine learning-based solution that forecasts revenue on a quarterly basis, including individual forecasts for 30 products in three different business segments. Specifically, the machine learning forecast used an elastic net, a random forest, a K-nearest-neighbor and a support vector machine. The winner model was then selected via back-testing. The forecasts generated in this way proved to be more accurate than the traditional forecasts generated by FP\&A in approximately 70\% of the cases. The paper cites the ability to incorporate external information (e.g., temperature as a driver for electricity demand) in regression frameworks  as an advantage of these over pure (standard) time-series models. While classical time-series are good at capturing trends and seasonality, they often struggle to incorporate external data. In particular, they generally lack a regularization mechanism. As a result, they tend to overfit the training data in high-dimensional settings, leading to low out-of-sample accuracy for new forecasts. Many machine learning methods include by design mechanisms to avoid overfitting (e.g., regularization for ridge, lasso and elastic net).\\

\cite{Barker2018} also highlight some requirements that arise from the intent to use the results of machine learning forecasts in a practical manner in a corporate setting. Traditionally, FP\&A works with a mid-point estimate, coupled with an estimation of the risks and opportunities around this mid-point. Risks and opportunities typically consist of a list of items or events that will materially impact the business results if they do not turn out as assumed in the mid-point forecast \cite{Conine2017}. Judgmental probability estimates provided by subject matter experts are often attached to these items, together with a quantification of the expected impact under the different scenarios.\footnote{Other methods with a very similar intent exist. Examples are the quantification of a best and worst case in addition to the normal or base case, or sensitivity analysis with varying degrees of sophistication.} For forecasts generated by traditional statistical or machine learning models, prediction intervals are therefore an important element for FP\&A practitioners in order to quantify the risk in the forecast. However, prediction intervals are not typically part of machine learning models. The solution proposed by \cite{Barker2018} consists of creating intervals from out-of-sample error distributions obtained during backtesting. Other practical requirements in a corporate environment are the need for a mostly automated solution allowing for fast forecast generation as well as the need to ensure high security standards for data storage, processing and access. Financial data such as sales and profits are highly sensitive and companies are reluctant to release them into public cloud environments. \cite{Barker2018} explain the details of their workflow automation and security controls, which revolve around the Microsoft Azure cloud-computing platform.\\

The third publication \cite{Koenecke2020} evaluated deep neural networks traditionally used in natural language processing (encoder-decoder LSTMs) and computer vision (dilated convolutional neural networks) in order to forecast company revenues. The approach incorporated transfer and curriculum learning. For the products and time period under study in this publication, deep neural networks improved predictive accuracy compared to the company's internal baseline, which combined traditional statistical and machine learning methods other than deep neural networks.\\

In another example of applied machine learning in the area of FP\&A, Daimler Mobility used an undisclosed library of machine learning algorithms to generate a monthly forecast set, spanning the next 18 months and updated monthly \cite{Unger2019}. In this respect, the approach followed the concept of a rolling forecast. The forecasted set of values comprised key financial performance indicators that were representative of Daimler Mobility's car rental, leasing, financing and fleet management business. According to \cite{Unger2019}, one of the key advantages of this approach compared to the traditional way of forecasting and budgeting is the speed with which updated forecasts are available, allowing faster adoption of corrective action.\\

These papers all discuss modern machine learning methods for financial forecasting. In the next section, we will show that these approaches cannot be applied directly to inference problems and how the double machine learning framework overcomes this problem. A first example will illustrate the use of machine learning techniques in FP\&A for forecasting. A second example will serve to illustrate the use of double machine learning for planning (inference). Finally, we will explore whether having additional data improves the results for both the forecasting and planning tasks.\\

\section{Simulation example}
\label{Simulation}
In this section, we provide the results of a small simulation study. The design of the simulation reflects the setting, types of data and questions that the FP\&A department in a large, multinational company could face. We will start with an example in which FP\&A is predominantly interested in the accuracy of sales forecasting. We will then carry this example forward into a question related to planning. In this second example, FP\&A is interested in assessing the effectiveness of promotional activities in generating sales; in other words, the question of interest relates to causal inference and the answer to this question can inform decisions about resource planning. Finally, we will investigate how the results change if the FP\&A department obtains additional data points for their tasks.

\subsection{Forecasting}

Assume for our stylized simulation the following setting:\footnote{We have intentionally kept the simulation example simple. For instance, we have not added any time-series-specific effects such as a trend component or serially correlated error terms. This allows us to focus on the key elements. For instance, the $N=60$  data points could represent observations in different countries or sub-markets, which would warrant a cross-sectional approach to the analysis. The conclusions presented in the simulation example will remain largely unchanged.} for a given month $n$, the FP\&A department would like to forecast the sales $y_{n}$ of a specific product or service. FP\&A has collected monthly data over five years ($N=60$) for sales as well as a set of $P=40$ factors or features that FP\&A believes could be predictive of these sales. We represent these factors as $x_{p,n}$ and the corresponding sales with $y_{n}$.\footnote{This is a case of supervised learning, because we have observations for both the input ($x_{p,n}$) and the output ($y_{n}$)} In practice, there can be a wide range of factors depending on the product or service. Examples include weather conditions and various macroeconomic indicators, but also specific customer shipment patterns or the current competitive market situation. Note that the size of the feature set can easily reach 40 plausible predictors once an initial, smaller feature set is increased due to the inclusion of transformed and newly created features. This step, called feature engineering, can include the creation of lagged variables (e.g., when the effect of the economic situation affects sales several months later) or interaction effects (e.g., when a particular weather situation coincides with a peak shipment date, nullifying or exacerbating the effect of the peak shipment date). A further example is the transformation of categorical variables into several binary values via so-called one-hot encoding (e.g., when classifying the competitive market situation as ``highly competitive'', ``moderately competitive'', ``not competitive'' and the like).\\

In addition to developing a set of 40 features, FP\&A measures the promotional activity carried out by the company for the product under investigation during the reference timeframe. We denote this promotional activity as $d_{n}$. For the purpose of this illustrative simulation, we work with the assumption that the promotional activity can be measured using a single variable. In other words, we do not enter into promotional mix considerations with interaction effects among the different promotional tools. In practice, this single variable could be a summary measure such as the amount of money spent on promotion and advertising; another possibility for a summary measure could be the number of customer calls or minutes of customer interaction. For forecasting and planning activities performed by FP\&A at aggregate levels, such as the regional, divisional or group levels, an approach like this, based on a summary measure, is sometimes applied. Extending the analysis to include several marketing variables is possible without any major changes.\\

Given the nature and intent of promotional activity, it appears natural for FP\&A to include $d_{n}$ in the list of likely predictors for the sales forecasting model. Furthermore, estimating the effect of the promotional activity on sales represents an important question for FP\&A, which we will address in the second part of this section, dedicated to planning.\\

In order to evaluate the accuracy of the sales forecasts, we will follow an out of sample evaluation approach. Only the first four years (48 data points) are used to build and train the forecasting models. FP\&A then compares the forecasts generated by the models to the actual values from the last year in the available dataset (12 data points). Note that these 12 data points have been intentionally excluded from the model creation phase. While more sophisticated training and evaluation strategies exist (e.g., rolling evaluation windows), the described approach is sufficient for the purpose of this simulation study because the out-of-sample forecasting performance is evaluated separately for each simulation.\\

For our simulation, we generate the data as n=1, \ldots, N independent and identically distributed (i.i.d.) draws from the following model:\\

\begin{align}
y_{n}=\alpha d_{n}+x'_{n}\beta_{p} +\varepsilon_{n}
\end{align}

and

\begin{align}
d_{n}=x'_{n}\gamma_{p} +\nu_{n},
\end{align}

with $x\sim\mathcal{N}(0,\Sigma)$ where $\Sigma$ is a p $\times$ p matrix with $\Sigma_{k,j}=c^{\lvert j-k \rvert}$,\\
$\varepsilon\sim\mathcal{N}(0,2)$ and $\nu\sim\mathcal{N}(0,2)$.\\

The second equation captures confounding, i.e., variables that are simultaneously correlated with the outcome variable and the variable of interest. By setting $\alpha = 0$, we assume that the promotional activity undertaken by the company has no effect on sales, i.e., that the promotion efforts are, in reality, a waste of resources. With $c = 0.3$ we include some moderate correlation\footnote{The value of c represents the correlation between immediate neighbor features (e.g., feature $x_{p}$ and feature $x_{p+1}$). Due to the way $\Sigma$ is constructed, the correlation decays quickly as the distance between features increases (e.g., feature $x_{p}$ and $x_{p+3}$ have only a correlation of $c^{3}$, which is 0.027 for c=0.3)} between features, which can be expected if several features from the same general background (e.g., macroeconomic factors) are included in the model.\\ 

We set $\beta = 0$, except for $\beta_{39}$ and $\beta_{40}$, both of which we set equal to 1. Thus, out of the 40 features included in the analysis, only two are actually related to sales. Similarly, we set $\gamma = 0$, except for $\gamma_{39}$ and $\gamma_{40}$, both of which we also set equal to 1. The two features related to sales also determine the amount of promotional activity $d_{n}$.\footnote{A simple example can help clarify the intuition behind this setting. Ice cream sales on the beach are probably positively related to weather conditions (feature 1) and the day of the week (feature 2). At the same time, the ice cream salesperson may decide to run some promotional activity when weather conditions are favorable on a weekend. Thus, the same features have an influence both on sales and on promotional activity. We will revert to this illustrative example in the section on planning.} $\varepsilon$ and $\nu$ are random error terms (so-called noise). We report results based on $1,000$ simulation replications.\\

It is important to remind ourselves that the FP\&A department naturally does not know any details about this data generation process. Only an oracle would know that, in reality, solely two of the 40 plausible predictors are linked to sales and that the coefficient in the data generating process is zero for the other 38 features. This situation characterizes sparse models. In such models, only a small number of many potential predictors and/or control variables are actually relevant \cite{Belloni2014a}. Identifying them leads to a correct model specification and is the main challenge.\footnote{By construction, our simulation example is exactly sparse, with parameter values for all non-relevant features exactly equal to zero. For practical applications, a more realistic assumption is approximate sparsity, meaning that all or many features can have non-zero parameter values. Nevertheless, only a limited number of features are needed to approximate the true relationship with sufficient accuracy. We refer interested readers to \cite{Belloni2010}. Our simulation could easily be extended to such a setting. Results would remain largely unchanged.} Additionally, FP\&A does not know that the promotional activity $d_{n}$ is correlated with the two features that have non-zero coefficients with respect to sales and that the promotional activity has no influence on sales ($\alpha$, the parameter of interest, is zero). We will come back to this point when we discuss inference.\\

We now provide results for two forecasting approaches. Both have in common that they rely - in this case correctly - on the typical assumption of a linear relationship between the output variable $Y$ (sales) and the full set of regressors $X$, which includes 40 presumably predictive features and one variable reflecting promotional activity.

\begin{align}
Y=X'\beta +\varepsilon,
\end{align}

The first approach is a traditional linear regression based on the ordinary least squares (OLS) method. Formally, OLS optimizes the parameters in such a way as to minimize the mean squared error (MSE):\\

\begin{align}
\hat{\beta}_{OLS} = \arg\min_{\hat{\beta} \in {\mathbb R}^p} \sum_{i=1}^n \{y_i -x'_i\hat{\beta}\}^2,
\end{align}

where $x'_i\hat{\beta}$ corresponds to the predicted sales value.\\

The second approach, post-lasso, is a classic machine learning technique. To estimate the coefficients, lasso uses a regularization strategy that is suited to high-dimensional problems in which the number of predictors exceeds or approaches the number of observations, as is the case in our simulation. In the first step, the lasso regression is performed. In the second (i.e., post-lasso) step, the method fits OLS on the coefficients selected in the first step. Formally, lasso optimizes the parameters to minimize MSE subject to a penalty for using parameters:\\ 

\begin{align}
\hat{\beta}_{Lasso} = \arg\min_{\hat{\beta} \in \mathbb{R}^p} \sum_{i=1}^n \{y_i -x_i'\hat{\beta}\}^2+\lambda\lvert \beta \rvert,
\end{align}

where $x'_i\hat{\beta}$ corresponds again to the predicted sales value.\\

The key difference between the lasso and OLS is that lasso minimizes a penalized MSE, in which the penalty amount corresponds to the absolute amount of each parameter included in the model, scaled by the tuning- or hyperparameter $\lambda$:

\begin{align}
Penalty_{Lasso} = \lambda\lvert \beta \rvert.
\end{align}

A detailed discussion of the theory behind regularization approaches would go beyond the scope of this article. Readers are referred, among many possible sources, to \cite{Hastie2009}, \cite{Buehlmann2011} or \cite{Taddy2019}. \cite{Taddy2019} sees regularization as ``the key to modern statistics'' by virtue of its ability to prevent overfitting in high-dimensional settings. Instead, we will recall a few characteristics of the lasso that are particularly relevant to our FP\&A example and the corresponding simulation.\\

The full name of the lasso (``least absolute shrinkage and selection operator'') indicates two important characteristics. First, as we can see in the formula for $Penalty_{Lasso}$, the absolute size of the coefficients included in the model represents a cost in the minimization of the MSE. Lasso will therefore shrink the coefficients towards zero. This makes the prediction system more stable and avoids overfitting. Second, the lasso-specific penalty in the form of the absolute value of the coefficients has the property that some parameters will be exactly equal to zero. In other words, the lasso will fully exclude some variables from the model and therefore perform automatic variable selection.\\

As indicated above, the lasso can handle situations in which the number of predictors approaches or even exceeds the number of observations. In our case, the number of predictors (including the measure of promotional activity) is 41 and the number of observations 48. Although OLS can still be calculated, we will see that its out-of-sample predictive accuracy becomes extremely unreliable. If we were to chose a simulation scenario with 48 or more predictors, OLS could no longer be computed. A second challenge for OLS in settings with many predictors is the increased risk of correlation among the predictors. If predictors are highly correlated among themselves, or if, in an extreme case, there is an exact linear relationship between two predictors (multicollinearity), OLS estimates become unstable. For instance, macroeconomic variables tend to be strongly correlated.\\

An important ingredient in the lasso is the size of the penalty, which depends on the tuning parameter $\lambda$. $\lambda$ is not determined by the lasso itself, but needs to be selected. Intuitively, $\lambda$ plays a role in filtering the relevant variables. Several strategies to select $\lambda$ have been proposed in the literature and are used by practitioners. The most common are cross-validation strategies and information criteria such as Akaike's or Bayes' information criterion. Our simulation study uses the data-dependent penalty level proposed by \cite{Belloni2013}. We refer interested readers to this source for details.\\

Compared to the standard lasso approach, which induces bias due to the shrinkage of coefficients, post-lasso has the advantage of a smaller bias, even if the model selected in the first step by lasso fails to include some of the true predictors. It also converges at a faster rate towards the true parameter values if the model selected by lasso correctly includes all true predictors (in addition to some irrelevant predictors). If lasso selects exactly (only) the true predictors, the post-lasso coefficient estimators are equal to the ones produced by an oracle that is aware of the underlying data-generating process \cite{Belloni2012, Belloni2014a}.\\

Table 1 summarizes the results of $1,000$ simulation runs for the forecasting task, comparing OLS to post-lasso.\\

\begin{longtable}{ccc}
  \toprule
RMSE & In sample & Out of sample  \\ 
  \midrule
OLS & 0.738 & 5.321  \\  
Post-lasso & 1.991 & 2.162  \\ 
   \midrule
 \bottomrule

\caption{Average RMSE of the hold-out-sample from $1,000$ simulation runs for the forecasting task}
\end{longtable}

We report the forecast accuracy in terms of average root mean squared error (RMSE) over all simulation runs both on the in-sample and the out-of-sample data set. As outlined above, the in-sample data set consists of 48 data points, which are used to build and train the models. The out-of-sample data set consists of 12 data points, which are intentionally not used in the model construction (``hold-out sample''), allowing the model to be evaluated on new, previously unseen data. The strong focus on forecasting performance on previously unseen data is a hallmark of the machine learning approach.\\

On the in-sample data, OLS produces a higher predictive accuracy than post-lasso, with an RMSE of 0.738 which is nearly one-third that of the post-lasso RMSE of 1.991. However, the real interest of the FP\&A department here is not to model past sales data. Rather, the predictive performance on new data is what matters to FP\&A; this is why the out-of-sample data have been set aside. Here, the OLS RMSE increases substantially to 5.321, more than twice as high as the post-lasso RMSE of 2.162.\\

We can draw two main conclusions from the simulation. First, the RMSE of standard OLS increases significantly between in-sample and out-of-sample data. This is because we have nearly as many features (regressors) as we have observations in the model. This leads to overfitting, whereby OLS uses the wealth of available features to adapt closely to the training data without searching for a potentially simple or sparse true underlying relationship. This yields a false sense of precision, which is immediately exposed when OLS is evaluated using previously unseen data. Second, the post-lasso RMSE is relatively stable between the in-sample and out-of-sample data. The in-sample performance is thus already indicative of the true predictive power when post-lasso is used on unseen data. Lasso achieves this through the regularization strategy described above, which leads to a very selective inclusion of features and thus parsimonious models. For reference, of the 40 available features in the simulation, post-lasso retains an average of only 1.2 as relevant and shrinks the coefficients of all the others to exactly zero. As a reminder, our simulation includes only two truly relevant features. The out-of-sample RMSE for post-lasso is thus slightly hihger than the perfect RMSE score of 2.0 (equal to the standard error that was selected for the noise parameter $\varepsilon$), which would be achieved by an oracle.\\

Figure 1 shows the distribution of the out-of-sample RMSE for the post-lasso forecast over the $1,000$ simulation runs. The distribution of the errors follows approximately a normal distribution (overlaid as a red line). From a practical perspective, the risk of generating a highly incorrect lasso forecast is therefore limited. Furthermore, the right tail of the lasso errors ends before the mean of the OLS error. This provides additional reassurance when relying on lasso. 

\begin{figure}[htbp]
	\centering
		\includegraphics[width=1.00\textwidth]{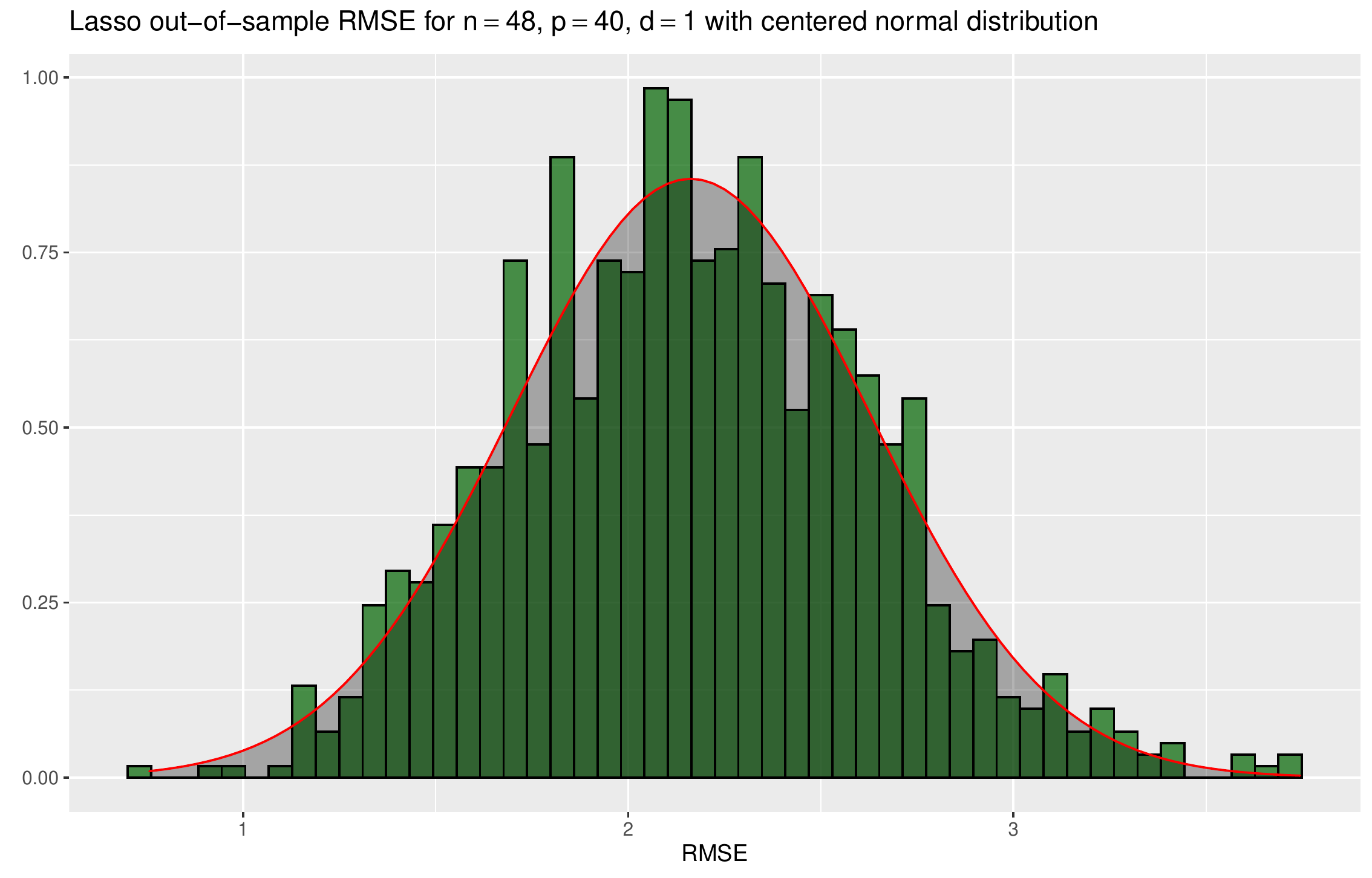}
	\caption{Figure 1: Distribution of the out-of-sample RMSE for the post-lasso forecast (bars), compared to the normal distribution (red line)}
	\label{fig:OOS_RMSE_Lasso}
\end{figure}

\subsection{Planning}

We will now discuss the use of machine learning in financial planning. To come back to our example, the task for the FP\&A department consists of evaluating the effectiveness of promotional activity in generating sales; in statistical parlance, the task relates to statistical inference of the effect of a treatment or intervention (i.e., the promotional activity) on an outcome (i.e., sales). This estimate forms the basis for planning and optimizing marketing activities. In our simulation examples, evaluating the effectiveness of promotion equates to estimating the parameter $\alpha$. As the parameter of interest, $\alpha$ corresponds to the effect of the promotional activity on sales, also called the ``lift'' in business applications. Let us remind ourselves that in our simulation, only two features are relevant for the sales forecast and that these two features also determine the amount of promotional activity. Thus, we are dealing with confounders because these two features are correlated with both the treatment and the outcome. Moreover, we have set $\alpha$ to zero, which effectively means that the promotional activity does not have an impact on sales.\\

In a business environment, this setting could correspond to an ice cream vendor at the beach who spends money on promotion whenever the weather is warm and sunny on the weekends. He ascribes the increased ice cream sales, or at least a part of them, to his promotional efforts, whereas in reality it is the favorable weather on the weekend that makes people come to the beach and enjoy his ice cream. Similar to the forecasting exercise, the FP\&A department is obviously not aware of the data generating process governing the simulation and needs to find a way to estimate $\alpha$.\\

One approach to estimating the effect of promotion could be to use the parameter estimate for $\alpha$ from the lasso model employed in sales forecasting. However, lasso shrinks parameter estimates because of the penalty loading used in the regularization process and therefore does not generate unbiased estimates of the parameter values, even though it allocates the least possible penalty amount to large signals while retaining the stable behavior of a convex penalty \cite{Taddy2019}. Additionally, lasso estimates predictors sparingly insofar as it sets many parameter estimates to exactly zero. In many cases, the factor measuring promotional activity ``may not make it'' into the second step of the post-lasso procedure.  It is therefore not meaningful to infer from the forecasting model the effectiveness of the promotional activities. We have previously highlighted the warning by \cite{Mullainathan2017} and \cite{Athey2018} that using a tool built for forecasting and assuming that its parameters possess the properties required for inference can be misleading.\\
 
With the above in mind, one could decide to pursue a hybrid solution with the following approach. Because the lasso has identified the most relevant features for prediction, we carry these forward into the inference model. Additionally, we include in the model the variable of interest (the intervention), which in our example is the variable that represents promotional activity. In a sense, we force this variable of interest into the model. We then estimate the parameter values for all of these features using OLS, which allows us to perform inference on the parameter estimates. In particular, we are able to interpret our parameter of interest $\alpha$ in this model. In our example, $\alpha$ will tell us the effectiveness of the promotional activities. Intuitively, a model that is constructed in this way can be understood as attempting to estimate the effect of promotional activity, while controlling for other factors with proven high predictive power from the forecasting model. For the ice cream vendor at the beach, this corresponds to controlling for the effect of the favorable weather during the weekend and thus deriving an isolated estimate of the effect of promotional activity on sales.\\

We will see from the simulation results that this approach, which we will call ``naive'', grossly fails to discover the true value of the parameter of interest; still, it is widely used by practitioners and applied researchers. In our model, the promotional activity measure is correlated with the features that concomitantly and directly influence sales. In the presence of such confounders, the naive approach will fail if the model selected for the forecasting task and serving as a basis for this naive approach is imperfect insofar as it omits some components of the true relationship between the feature set and the outcome variable. In high-dimensional settings, which require regularization methods such as the lasso, perfect model selection is an unlikely scenario and holds only under unrealistic assumptions. In short, the naive approach will suffer from omitted variable bias. This is because machine learning methods capture the features correlated with the outcome variable and deliver good predictive performance but often miss variables that are correlated weakly with the outcome but correlated more strongly with the intervention variable. Missing these variables does not harm predictive performance but biases the estimation of the intervention effect, leading to invalid post-selection inference.\\

For an approach to be valid, it must overcome this problem of imperfect model selection. Double or debiased machine learning, as proposed by \cite{Chernozhukov2017}, is one way to do so. The fundamental idea\footnote{Double machine learning also uses cross-fitting, an efficient way of data splitting. Interested readers are referred to \cite{Chernozhukov2017}.} is to reduce, for the estimation of the parameter of interest (i.e., the intervention variable), the sensitivity with respect to errors in selecting and estimating the nuisance parameters (i.e., the other predictors in the model). Technically, this can be achieved by regressing residuals on residuals. The first set of residuals is generated by regressing the outcome variable on the control features, notably using regularizing machine learning methods such as (post-)lasso, random forests, boosted trees or other methods suited for high-dimensional settings. The second set of residuals is generated by regressing the treatment variable on the control features. This auxiliary step helps to find the confounders that might lead to omitted variable bias. Finally, the first set of residuals is regressed on the second set of residuals. The parameter value obtained in this residuals-on-residuals regression represents the effect of the treatment variable on the outcome. Translated into our stylized simulation, the first regression relates sales to the 40 presumably predictive features; the differences between the predictions from this first regression and the actual outcomes (sales) constitute the first set of residuals. The second regression relates the promotional activity score to the 40 presumably predictive features; the differences between the predictions from this second regression and the actual outcomes (promotional activity) constitute the second set of residuals. Concretely, we will use a post-lasso approach for both cases, but in principle any other machine learning method could be used. Finally, we regress the residuals from the first regression onto the residuals from the second regression to obtain an estimate for the parameter of interest, $\alpha$, which represents the impact of promotional activities on sales.\\

This approach works well in practice \cite{Athey2018}, because the residuals-on-residuals approach makes the estimation of the treatment effect immune (i.e., orthogonal) to errors in the model specification. This is why the family of approaches that use this principle is also referred to as orthogonal machine learning \cite{Taddy2019}. Interested readers are referred to the literature for an in-depth theoretical discussion, including underlying assumptions and formal proofs, which is beyond the scope of this paper. Key sources include \cite{Belloni2014a}, \cite{Chernozhukov2015} and \cite{Chernozhukov2018}. To implement our simulation, we use the partialling-out approach as defined by \cite{Chernozhukov2016a} and report the corresponding results under this label.\\

Table 2 summarizes the results of the two approaches (i.e., ``naive'' and ``partialling out'') from $1,000$ simulation runs. We report the mean estimate for $\alpha$, the standard deviation of the estimate and the corresponding t-statistic and p-value for a two-sided test of whether the mean is different from zero. The rejection rate represents the proportion of individual simulation runs in which the ingoing assumption of $\alpha$=0 has been rejected based on the t-test (at the customary 5\% significance level). In other words, these are the instances in which the model incorrectly suggests an effect (positive or negative) of promotional activity on sales.\\

\begin{longtable}{ccc}
  \toprule
$\alpha$ & Naive & Partialling-out  \\ 
  \midrule
Mean estimate & 0.1604 & 0.0081  \\  
Std. dev. & 0.1668 & 0.1326  \\ 
t-statistic & 30.422 & 1.934  \\ 
p-value & 0.0000 & 0.0534  \\ 
Rejection rate & 46.1\% & 4.8\%  \\ 
   \midrule
 \bottomrule

\caption{Results from 1000 simulation runs for the planning/inference task. The t-statistic and p-value refer to the respective mean estimate.}
\end{longtable}

The simulation results provide several insights. First, and this is the main point we seek to make, the naive approach grossly fails to discover the true value of $\alpha$, because it suffers from significant bias. Put simply in the context of our simulation, this bias represents systematic over-estimation of $\alpha$ and thus over-estimation of the effectiveness of promotion. On average, the naive approach estimates a value for $\alpha$ of 0.1604, compared to a true value of zero. The partialling-out approach also yields an average positive value for $\alpha$ of 0.0081, but is much closer to the true value of zero. Relatively speaking, the bias of the naive approach is roughly 20 times higher than that of the partialling-out approach.\\

A second point is that the standard deviation of the estimates for $\alpha$ are similar for both approaches. Figure 2 (naive approach) and figure 3 (partialling-out approach) show the distribution of the estimates for $\alpha$ from the $1,000$ simulation runs compared to a normal distribution curve. Visual inspection suggests that the shapes of both distributions are well-approximated by a normal distribution. Of course, the center of the distribution for the naive approach is clearly shifted to the right of zero. This reinforces the point made above that bias is induced by the naive approach.\\

\begin{figure}[htbp]
	\centering
		\includegraphics[width=1.00\textwidth]{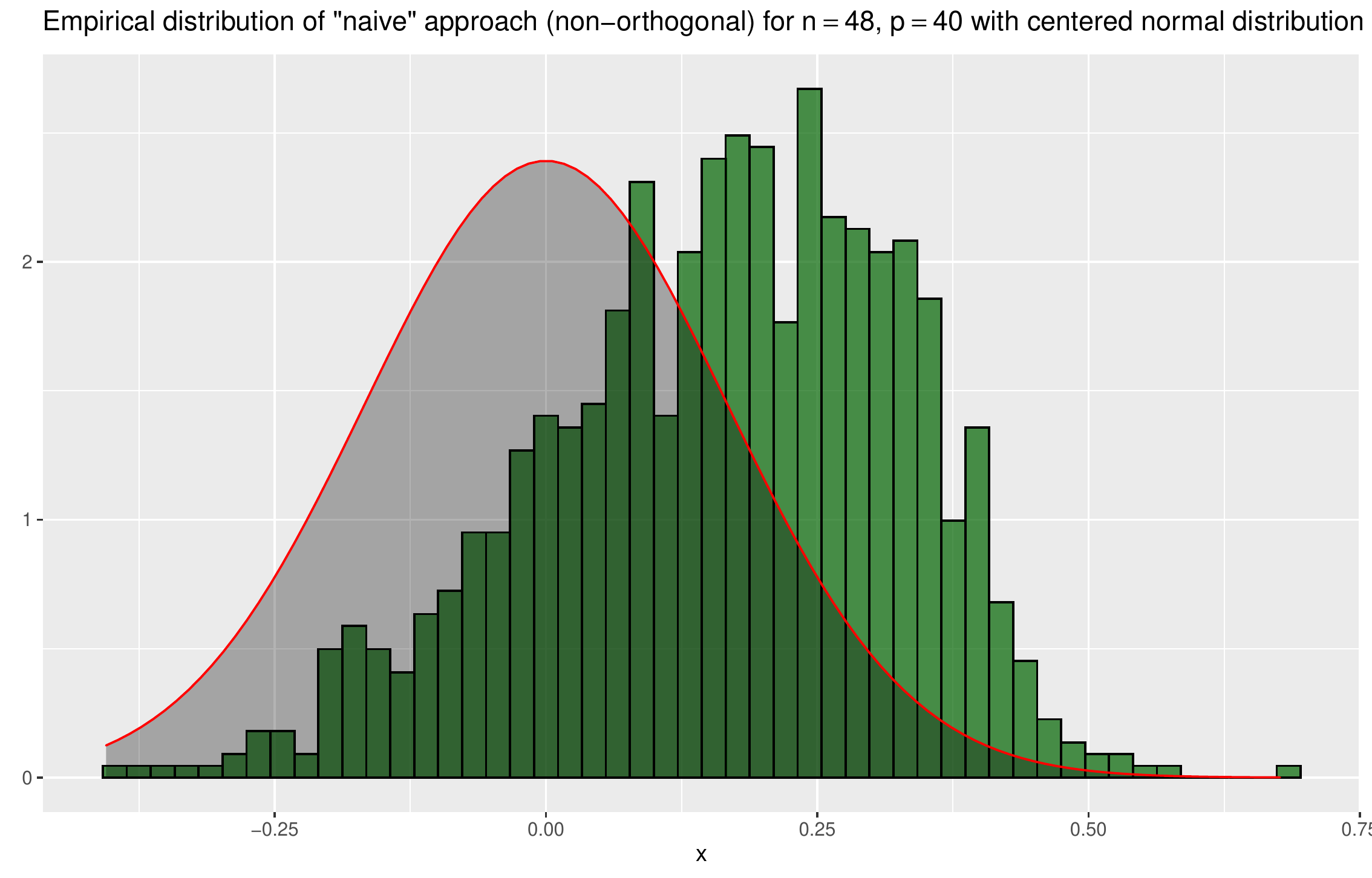}
	\caption{Distribution of estimator for $\alpha$ from the naive approach}
	\label{fig:Alpha_Naive}
\end{figure}

\begin{figure}[htbp]
	\centering
		\includegraphics[width=1.00\textwidth]{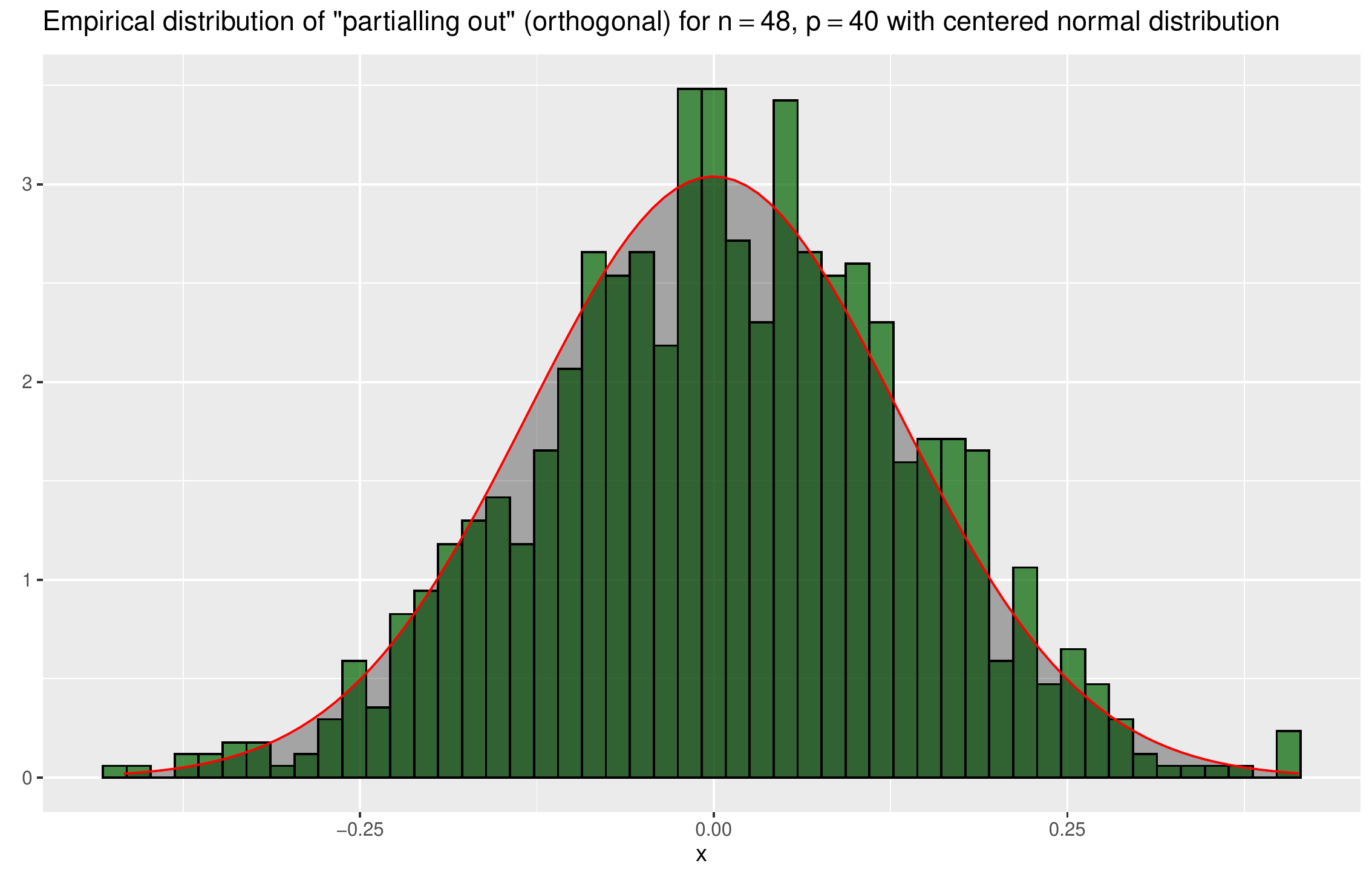}
	\caption{Distribution of estimator for $\alpha$ from the partialling-out approach}
	\label{fig:Alpha_partial}
\end{figure}

Table 2 also reports the t-statistic and corresponding p-value for a two-sided test of whether the mean estimate of $\alpha$ is zero. Under the naive approach, this hypothesis would be rejected with high confidence (t-statistic of 30), reinforcing the incorrect belief that the promotional efforts positively affect sales. Under the partialling-out approach, the hypothesis of no effect from promotional efforts would not be rejected at the customary 5\% threshold level (t-statistic of 1.93). In practice, the FP\&A department would of course not benefit from this kind of insight as they would not have access to repeated estimates for $\alpha$. With the advantage of being able to run multiple simulations, we can use this information to support the point of significant bias in the naive approach. Nevertheless, the rejection rate, reported in the last line of table 2, provides a good indication of how often FP\&A would make an incorrect decision. For each individual run in the simulation, this metric records whether FP\&A would (incorrectly) reject the assumption that $\alpha$ is zero at the typical 5\% significance level. Under the naive approach, this would happen 46\% of the time. Put differently, a bit less than half of the time, FP\&A would incorrectly assume that promotional activity does have an effect on sales. With partialling-out, this error drops to slightly below 5\%.\footnote{Note that one would expect an error rate here of around 5\% from a correct model because the 5\% significance level corresponds to a 5\% probability of rejecting the null hypothesis when it is, in reality, true.} \\
   
In summary, by relying on the naive approach, the FP\&A department (or the ice cream vendor) would substantially overestimate the causal effect of the promotional activity on sales. Consequently, this activity would probably be maintained or even increased for this product or service, even though in reality, it does not increase sales. Put differently, the company would draw up plans that allocate resources wastefully on this particular product or market. The impact from falling into this trap could multiply even further across the organization if the results of such an analysis were used as a benchmark for similar products, services or geographic markets. This might happen, for example, if data are not readily available for a particular product (for example, one that is being newly launched) and the decision is made to extrapolate from existing (and potentially wrong) information. Such a situation is even more likely when existing information appears plausible and suitable\footnote{\cite{Blake2015} highlight in their paper that ``[...] the incentives faced by advertising firms, publishers, analytics consulting firms, and even marketing executives within companies, are all aligned with increasing advertising budgets.''} and, in addition, is perceived as objective, unbiased (in the sense of free from human/cognitive bias) or even scientific because it was generated using data-driven methods.\\

\subsection{The value of data}
In 2017, ``The Economist'' \cite{Economist2017} asserted in the title of its May 6 edition that data are now the world's most valuable resource. Questions about the value of data as a resource and production factor have generated great interest in academia and policy institutes. One consideration within this vast topic is a (hypothesized) positive feedback loop: more data lead to more data-driven insights, allowing a company to serve its customers better, to attract more customers and, in turn, to collect even more data. Nevertheless, there seems to be a broad consensus that data are generally governed by decreasing returns to scale, like any other production factor \cite{Varian2018}, \cite{Bajari2019}.\\

In this paper, we will limit ourselves to a short discussion of how the number of observations affects the accuracy achieved by the forecasting and inference methods used by the FP\&A department within the frame of our simulation. For empirical results, we refer interested readers to \cite{Bajari2019}, which contains a study of the performance of Amazon's retail forecasting system. The study finds performance gains in the time dimension (i.e., from longer data history), but not in the product dimension (i.e., panel data forecasts do not improve with more products within a category). An interesting finding is the overall improvement of forecasts over time (controlling for the length of data history and the number of products), suggesting positive effects from improved technology (e.g., new machine learning models, better hardware or adaptation of organizational practices). \\

In our simulation, the hypothetical FP\&A department uses a training set of 48 observations, 40 predictive features and one variable of interest for inference (i.e., the measure of promotional activity). In many real-life applications relevant to FP\&A departments, the number of observations available for analysis is typically limited. More observations may simply not exist; for instance, new products generate sales data starting only from their launch date. Even if data do exist, collecting, accessing and, if necessary, curating them come at a cost; for instance, companies may limit the amount of directly accessible data history due to system constraints, or data generated prior to the introduction of new software may be inaccessible, in full or in part.\\

Let us now explore simulation results assuming that the FP\&A department has invested in expanding the set of observations to 72. The number of features (i.e., 40), the variable of interest (promotional activity measure) and the overall simulation set-up remain unchanged.\footnote{We intentionally do not allow the number of features to grow with the sample size (see for instance \cite{Belloni2010}) in order to isolate the effect of the additional observations clearly. In practice, a significant extension of the number of observations may require including additional control features.} We again run $1,000$ simulations. What is the return on accuracy of expanding the observation set?\footnote{The natural way to think about the expansion is to assume that the department ``digs out'' additional historical observations. However, from a theoretical standpoint, the department could also wait two years and gather the additional data points over time. In this case, the change in forecasting accuracy could be mistaken for a technology or learning effect by an outside observer.}\\

Table 3 reports the forecasting results based on 72 observations compared to the previous simulation based on 48 observations. For OLS, the in-sample accuracy drops, as witnessed by the increase in RMSE to 1.309 from the previous RMSE of 0.738 with 48 observations. However, the out-of-sample accuracy increases: the corresponding RMSE drops to 2.982 from the previous RMSE of 5.321. In fact, the additional observations reduce the extent of overfitting seen in the initial setting. With 40 features and (only) 48 observations, OLS was actually close to the point of failing. This point would have been reached if the number of features had been equal to or exceeded the number of observations. Intuitively, OLS moves further away from this point by expanding the set of observations (and keeping the number of features constant).\\
\\
\\
\\



\begin{table}[]
\begin{tabular}{lllll}
\hline \hline
           & \multicolumn{2}{l}{48 trainings observations} & \multicolumn{2}{l}{72 trainings observations} \\
\hline
RMSE       & In sample           & Out of sample           & In sample           & Out of sample           \\
\hline
OLS        & 0.738               & 5.321                   & 1.309               & 2.983                   \\
Post-Lasso & 1.991               & 2.162                   & 1.988               & 2.085      \\
\hline \hline
\end{tabular}

\caption{Average RMSE from $1,000$ simulation runs for the forecasting task, 48 vs. 72 training observations.}
\end{table}

For post-lasso, the results based on 72 observations are quite similar to those obtained with 48 observations. Neither the in-sample nor the out-of-sample RMSE change notably. As expected and unlike OLS, post-lasso already deals well  with the initial situation in which the number of features is close to the number of observations and benefits only marginally from the increase in observations. Put differently, post-lasso does not require investing in the generation or acquisition of additional data. Our finding is consistent with standard stochastic theory.\footnote{See for instance \cite{Buehlmann2011} or \cite{Belloni2013}. Post-lasso converges towards the true parameter value at a rate of $n^{-1/4}$, which is slower than the OLS rate of $n^{-1/2}$. The value of additional data is thus generally smaller for post-lasso than for OLS.}\\

In summary, while having more data is generally beneficial, expanding the observation set for forecasting in our simulation study creates a tangible advantage only for OLS. If the FP\&A department employs post-lasso, which is the preferable method in this setting, the gain in precision from expanding the observation set is very small and, for many practical applications, would not warrant the effort.\\

We will now look at inference, which entails estimating the (causal) effect of promotional activities on sales. Table 4 reports the inference results for $\alpha$ based on 72 observations compared to the previous simulation based on 48 observations. Recall that the true value of $\alpha$ is zero. For OLS, the mean estimate for $\alpha$ decreases to $0.0956$ (from the previous estimate of $0.1604$ with 48 observations). However, based on a standard t-test, this value is still significantly different from zero (t-statistic of $17.847$). For the partialling-out approach, the mean estimate for $\alpha$ decreases by a factor of 10 to $-0.0008$ (from 0.0081 with 48 observations). Based on a standard t-test, it is now virtually indistinguishable from zero (t-statistic of -0.217).\\



\begin{table}[]
\begin{tabular}{lllll}
\hline \hline
                      & \multicolumn{2}{l}{48 trainings observations} & \multicolumn{2}{l}{72 trainings observations} \\
\hline
$\alpha$ & Naive            & Partialling-out            & Naive            & Partialling out            \\
\hline
  Mean estimate & 0.1604 & 0.0081 & 0.0956 & -0.0008  \\  
Std. dev. & 0.1668 & 0.1326 & 0.1694 & 0.1166  \\ 
t-statistic & 30.422 & 1.934 & 17.847 & -0.217  \\ 
p-value & 0.0000 & 0.0534 & 0.0000 & 0.8279  \\ 
Rejection rate & 46.1\% & 4.8\% & 38.6\% & 6.4\%  \\  
\hline \hline
\end{tabular}
\caption{Results from $1,000$ simulation runs for the planning/inference task, 48 vs. 72 training observations. The t-statistics and p-values refer to the respective mean estimates.}
\end{table}

The expanded set of observations reduces the bias of the naive approach. Intuitively, the risk of imperfect model selection described above becomes smaller. Still, the naive approach exhibits significant bias compared to the true value of $\alpha$. For the partialling-out approach, the additional observations lead to a mean estimate for $\alpha$ that comes very close to the true value. Depending on the required precision of the estimate, the FP\&A department could benefit from the additional set of observations in its analysis. Again, our finding is consistent with standard theory on convergence rates (see for instance \cite{Buehlmann2011} or \cite{Belloni2013}). Whereas post-lasso converges for forecasting towards the true parameter value at a relatively slower rate of $n^{-1/4}$, the double machine learning estimator of the treatment effect converges at the faster rate of $n^{-1/2}$ (i.e. the same rate as OLS).


\section{Conclusion}

Digitalization, especially when it couples large amounts of data with appropriate tools for analysis, represents an important opportunity for the financial planning and analysis function. In this article, we have provided an introductory overview of machine learning in this context. By reviewing several relevant theoretical aspects of machine learning and discussing the results of a simulation study, we have demonstrated how machine learning may prove useful for FP\&A practitioners. We have paid special attention to explaining the distinction between forecasting and planning tasks, the first of which involves prediction and the latter of which involves causal inference. We see the confusion of these two concepts as a major pitfall that practitioners should strive to avoid. Specific approaches to causal machine learning have begun to gain traction, as awareness has increased that the naive application of machine learning can fail in applications that go beyond prediction. This applies to all modern machine learning methods in a high-dimensional setting.\\

Our article has several limitations. It was impossible to cover the vast number of machine learning techniques that exist. Depending on the causal question at hand, a range of econometric approaches (e.g., instrumental variables, synthetic controls or regression discontinuity designs) coupled with machine learning methods may be suitable. We intentionally used a simple data generation process in our simulation; additional elements such as trends or seasonal components or a real-life example could complement our simulation study. Despite these limitations, we believe that our article can be a valuable source of insights into the ways in which FP\&A can benefit from machine learning. With it, we hope to contribute to the adoption of machine learning in this area and help practitioners avoid common mistakes.


\newpage
\footnotesize
\pagebreak
\bibliographystyle{imsart-number}
\bibliography{Literatur_HW_v02_April21}

\end{document}